\documentclass[%
 reprint,
 amsmath,amssymb,
 aps,
]{revtex4-1}

\usepackage{graphicx}% Include figure files
\usepackage{dcolumn}% Align table columns on decimal point
\usepackage{bm}% bold math
\usepackage{hyperref}% add hypertext capabilities
\usepackage{braket}
\begin{document}

\title{An Introduction to the Inverse Quantum Bound State Problem in One Dimension}% Force line breaks with \\

\author{Thomas D. Gutierrez}
\email{tdgutier@calpoly.edu}
 \homepage{http://www.tdgutierrez.com}
\affiliation{
 Physics Department, California Polytechnic State University, San Luis Obispo, CA 93407
}%
\date{August 2013}

\begin{abstract}
A technique to reconstruct one-dimensional, reflectionless potentials and the associated quantum wave functions starting from a finite number of known energy spectra is discussed.  The method is demonstrated using spectra that scale like the lowest energy states of standard problems encountered in the undergraduate curriculum such as: the infinite square well, the simple harmonic oscillator, and the one-dimensional hydrogen atom.  

\end{abstract}

\maketitle
\section{\label{sec:intro}Introduction}
Undergraduate students in an introductory modern physics or quantum mechanics course are exposed to an array of mathematical methods that aid in solving the time-independent Schr\"odinger equation as an eigenvalue problem \cite{Harris07, Griffiths05}.  The topic is framed in the ``direct'' or ``forward'' sense: given some potential function, find the corresponding allowed energy eigenvalues and their associated eigenfunctions.
This problem might be set up either in the context of a bound state problem, where the resultant eigenvalues and functions form a discrete set, or a scattering problem, where the relative reflection and transmission coefficients can be calculated. Advanced quantum mechanics courses at the undergraduate and graduate level often cover the forward scattering problem in some detail, developing the mathematical methods of partial wave analysis, S-matrix theory, the Born approximation, and other standard analytical techniques \cite{Shankar94,Griffiths05,Sakurai10}. 

Advanced quantum mechanics courses sometimes introduce students to the inverse scattering problem: given information about both the the asymptotic initial and final states of a system, what information can be gleaned about the  interactions that facilitated this transition \cite{Chadan77}?  Although powerful, these kinds of problems can be nontrivial and often involve laborious analytical or computational efforts.  Nevertheless, the inverse quantum scattering problem forms the foundation of many contemporary lines of scientific and applied research.  For example, in modern nuclear and particle physics \cite{Chadan77,Mackintosh12,Halzen84, Griffiths08}, information about the nature of fundamental interactions between colliding systems and their constituents can be gleaned by observing the production and scattering rates of different kinds of final states.  In an applied setting, medical diagnostic tools such as PET scans and CT scans are able to reconstruct images based on emission tomography technology, which relies heavily on inverse scattering methods \cite{Wernick04}.

However, while the inverse scattering problem is occasionally covered in more advanced courses, the inverse problem for bound states is typically not.  The inverse problem for bound systems can be stated as follows: 
\begin{quote}
\textit{given a complete set of negative bound state energies as well as the reflection and transmission coefficients for all positive energies, find the associated potential and wave functions that solve the Schr\"odinger equation.}  
\end{quote}
Its absence in the introductory curriculum is not entirely surprising because the formal treatment of the problem in the literature \cite{Faddeyev63, Chadan77} does not readily lend itself to a first exposure to the topic.  Also, the utility of the inverse bound state problem is not as well established as the inverse scattering problem in the traditional quantum curriculum.  Yet, some discussion of the inverse bound state problem in an upper division quantum mechanics course is perhaps in order.  For example, although still experimentally challenging, technologically generating customized potentials to achieve desired energy spectra is becoming a possibility, such as the case with quantum dots, optical lattices, or nano heterostructures \cite{Bimberg99, Jessen96, Deutsch00, Nam13}.  For applied quantum bound state problems it is arguably, in many cases, the energy spectrum, not the shape of the potential, that is of technological or scientific interest.  For example, if an application aims to emit or absorb a photon of a particular wavelength or if a certain band structure is desired to affect the conductivity of a material, manipulating the energy levels of the system are of keen interest.

While the nuanced formalism underlying the inverse bound state problem may be too advanced for introductory courses, the resultant mathematical technology, once established, is not more difficult to implement than other one-dimensional standard bound state or scattering problems, as shown below.    

The exercise of exploring the relationship between potentials and their spectra has obvious pedagogical value, a relationship normally explored in the forward sense by modifying the potentials and seeing the effect on the spectra.  However, introducing the inverse problem provides a powerful pedagogical platform for students to study the relationship between the energy eigenstates and the potentials in a way not normally covered in standard quantum mechanics courses.  Moreover, solving the inverse bound state problem is engaging; it is intellectually satisfying to start with user-customized energy spectra and work backwards to find novel analytical forms for potentials.  

Finally, the methods make contact with some very deep ideas beyond just the practicality of performing an inverse transformation.  For example, inverse methods similar to that discussed in this work have been used to find potentials whose quantum energy spectra reproduce the prime number density and Riemann Zeta Function zeros \cite{Muss97, Schum08}, which in turn links quantum mechanics to principles in number theory.

While the formalism is quite general, the treatment below limits itself to a remarkable class of potentials known as {\em reflectionless potentials}, where the reflection coefficient is zero while the transmission is unity.  Reflectionless potentials focus the effort on the inverse bound state problem and make some of the mathematics more tractable. Because of the curious reflectionless nature of the potentials, there are natural connections with solitons \cite{Novikov84} and even supersymmetric quantum mecahnics \cite{Cooper95}.  An accessible analysis of scattering and bound state characteristics of a particular class of reflectionless potentials can be found in references \cite{Kiri98, Lekner07}.

In this work, without excessive formalism, students will be given the necessary tools to explore inverse bound state problems themselves.  After setting up the mathematical algorithms, a few examples to demonstrate the method are outlined.  Several exercises, with some hints and solutions, are also provided to generate ideas for instructors and students alike.

\section{\label{sec:inverse} The Inverse Bound State Problem}

Given the {\em scattering data} for a system, the bound state negative energies as well as the reflection and transmission coefficients for all positive energies, can a potential function and energy eigenfunctions be found that are consistent with the Schr\"odinger equation?  To address this problem, one approach amongst many \cite{Faddeyev63, Barcilon74, Chadan77, Novikov84} is developed here, which finds solutions to the Mar\u{c}henko integral equation, a special case of the Gel'fand-Levitan equation \cite{Chadan77, Novikov84}, which can be used as an integral representation of the Schr\"odinger equation.  Following Reference \cite{Schum08} this will be referred to as the ``Mar\u{c}henko method.''   A detailed proof of the Mar\u{c}henko and associated methods is related to the inverse Sturm-Liouville problem and the formal procedure for solving the inverse bound state problem is well established \cite{Kay56, Faddeyev63, Barcilon74, Chadan77, Novikov84}. However, it is beyond the scope or intention of this work to derive it from first principles or explain it in detail.  The objective is rather to provide a set of distilled mathematical tools and procedures the motivated undergraduate can use.  Readers are encouraged to further investigate the underlying formalism as contained in the references \cite{Kay56, Faddeyev63, Barcilon74, Chadan77, Novikov84, Schum08}.

The Mar\u{c}henko method outlined here is appropriate for symmetric one-dimensional potentials and incorporates both scattering and bound state properties of the system.  Armed with this information, the time-independent Schr\"{o}dinger equation can be inverted for the potential $V(x)$, the bound state wave functions $\psi_n(x)$, and the scattering wave functions, $\Psi(x,E)$.  Although an abuse of nomenclature, the term ``potential'' will be used for ``potential energy,'' a practice not inconsistent with the existing textbooks and literature.  It is always understood here that ``the potential'' is a potential energy.

With the physical constants expressed in convenient units where $\hbar^2/(2m)=1$, the Schr\"{o}dinger equation for the bound system can be expressed as
\begin{equation}
-\psi_n''(x)+V(x)\psi_n(x)=E_n\psi_n(x),
\label{se}
\end{equation}
where $\psi_n(x)$ is the $n$th energy eigenstate and $E_n$ is the associated $n$th energy eigenvalue.  In this work, $\psi_n(x)$ refers specifically to the bound state wave functions in contrast to the scattering wave functions introduced later in Eq.~\ref{scat}.  The potential is taken to be negative for all $x$ as are the bound state energies, $E_n$.  Primes indicate spatial derivatives with respect to position $x$.  The discrete index $n$ is included as a reminder that Eq.~\ref{se} is representing a set of discrete states.

\subsection{Obtaining the potential}
Let there be $N$ negative bound state energy eigenvalues parameterized by $E_n=-\kappa_n^2$ with $n=1, 2,...,N$ and real $\kappa_n$.  They are ordered such that $n=1$ is the ground state (most negative) energy so $\kappa_1>\kappa_2>...>\kappa_N$.  Also, let the potential be one-dimensional, symmetric about the origin, everywhere negative, and smoothly approaching zero from below at infinity.  As mentioned above, the potential emerging from the following formalism will also be reflectionless.  That is, the reflection coefficient is zero and so particles with all energies $E>0$ impinging on the potential are completely transmitted.  With these assertions, the scattering data for the problem have been completely defined.  Under these conditions, it can be shown \cite{Kay56, Novikov84} that the potential solving the Schr\"{o}dinger equation is of the form

\begin{equation}
V(x)=-2\frac{d^2}{dx^2}\ln{(\det{({\bf I}+{\bf C})})}
\label{mV}
\end{equation}
where ${\bf I}$ is an $N\times N$ identity matrix.  The elements of the matrix ${\bf C}$ are given by
\begin{equation}
C_{ij}=\frac{c_i c_j}{\kappa_i+\kappa_j}e^{-(\kappa_i+\kappa_j)x}
\label{CC}
\end{equation}
and the set of $N$ normalization constants, $c_n$, are generated by
\begin{equation}
c_n^2=2\kappa_n\prod_{m=1,m\ne n}^{N}\left|\frac{\kappa_m+\kappa_n}{\kappa_m-\kappa_n}\right|.
\label{cn}
\end{equation}

While the origins of these equations are not immediately obvious, the end result is actually straightforward and generates the analytical form for the potential given the energy eigenvalues.  

\subsection{Obtaining wave functions}
A nice extension of the Mar\u{c}henko method developed by Kay and Moses \cite{Kay56} is used to extract the associated bound state and scattering wave functions from a given finite set of $N$ known energy eigenvalues.   The time-independent Schr\"{o}dinger equation for the scattering states, where $E>0$ with $V(x)<0$ for all $x$, takes the form

\begin{equation}
-\Psi''(x,E)+V(x)\Psi(x,E)=E\Psi(x,E),
\label{scat}
\end{equation}
which looks similar to the bound state system Eq.~\ref{se}, except the scattering wave function, $\Psi(x,E)$, and energy, $E$, are no longer discrete.  However, under these conditions, the form of $\Psi(x,E)$ can be written in terms of the bound state eigensystem, $\kappa_n$ and $\psi_n(x)$, in the following form \cite{Kay56},

\begin{equation}
\Psi(x,E)=\left[1+\sum_{n=1}^N \frac{\psi_n(x)}{\kappa_n+i\sqrt{E}}e^{\kappa_n x}\right]e^{i\sqrt{E}x},
\label{scatwf}
\end{equation}
which is taken as a given starting point.  The $N$ energies $E_n=-\kappa_n^2$ and wave functions $\psi_n(x)$ separately satisfy the bound state problem for the potential $V(x)$ in Eq.~\ref{se}.  Although perhaps not instantly apparent, the form of Eq.~\ref{scatwf} can be appreciated by noting scattering solutions to the Schr\"odinger equation for reflectionless potentials have the asymptotic form $\sim e^{i\sqrt{E}x}$ as $x\rightarrow-\infty$ (incident) and $\sim e^{i\delta} e^{i\sqrt{E}x}$ as $x\rightarrow+\infty$ (transmitted) for real $\delta$;  the reflection term, $\sim e^{-i\sqrt{E}x}$, is conspicuously absent.
 
Also, taken here as another given, it can be shown that the $\psi_n(x)$ satisfy the following $N$ equations

\begin{equation}
c_n^2 e^{\kappa_n x}\sum_{\nu=1}^{N}\left(\frac{e^{\kappa_\nu x}}{\kappa_n+\kappa_\nu}\right)\psi_\nu(x)+\psi_n(x)+c_n^2 e^{\kappa_n x}=0
\label{km}
\end{equation}
where, as before, $n=1, 2,...,N$,  and $c_n$ are the normalization constants obtained from Eq.~\ref{cn} above. The fully normalized bound state wave functions are $\psi_n(x)/c_n$ (with the above formalism, the extracted $\psi_n(x)$ themselves are not normalized).

\subsection{Summary procedure}\label{sum}
In an effort to help students quickly get started on projects generating potentials and wave functions from their own energy spectra, the previous discussion can be distilled as follows:

\begin{enumerate}
\item Select a finite, ordered set of $N$ $\kappa_n$ related to the energy via $E_n=-\kappa_n^2$ ($\kappa_1$ is the ground state and so is the largest of the set with $\kappa_N$ being the smallest);
\item \label{cnstep}Generate the set of $N$ normalization constants, $c_n$, using Eq.~\ref{cn};
\item Generate the set of matrix elements for ${\bf C}$ using Eq.~\ref{CC};
\item Obtain the functional form of the reflectionless potential $V(x)$ using Eq.~\ref{mV} consistent with the initial energy eigenstates;
\item Use standard linear algebra with the set of $\kappa_n$ and $c_n$ to solve Eq.~\ref{km} for the $N$ bound state wave functions $\psi_n(x)$; normalize them by dividing by the corresponding $c_n$;
\item Use the unnormalized bound state wave functions, $\psi_n(x)$, to construct the scattering state wave functions for this reflectionless potential using Eq.~\ref{scatwf}.
\end{enumerate}

To obtain analytical solutions for any system with more than a couple of eigenstates, the procedure outlined is best followed using software capable of symbolic manipulation.  A disadvantage to this analytical approach is that, after a fairly modest number of eigenstates, the calculation becomes numerically hypersensitive to the precision.  Also, the algebra, even for a computer, becomes cumbersome and time consuming after six or more states.  Thus, for formal research applications constrained by a rich energy spectrum and requiring high precision, considerable computing power, or an entirely different numerical approach, may be necessary.  However, precise interactive analytical results can be generated quickly using straightforward code reliably for roughly 4 states.  The potential shape solutions tend to be more stable than the wave functions, and potentials that support 6 or 7 states can usually be found.  The wave functions tend to become numerically unstable after 4 or 5 states.  Although not ideal, this is certainly sufficient to provide students and instructors alike interesting pedagogical avenues to explore.

\subsection{Examples}
Here are some standard systems analyzed using the Mar\u{c}henko method to obtain the potentials.   The Kay and Moses method was used for obtaining the associated normalized wave functions.  In all cases, as discussed above, the potentials are reflectionless, but their curious shapes are crafted so their bound states match the first few energy levels of the associated sample system.  All potentials and energy levels are negative, consistent with the formalism developed.  The simple harmonic oscillator is discussed in \cite{Schum08} and further clever computationally intensive numerical methods are discussed to greatly extend the number of states that can be explored.  The other potentials described below were generated for this paper and have not been treated in the literature using the methods described.   The ``a.u.'' in the plots stands for ``arbitrary units'' with $\hbar^2/2m=1$.  

\begin{figure}[h!]
\resizebox{.5\textwidth}{!}{\includegraphics{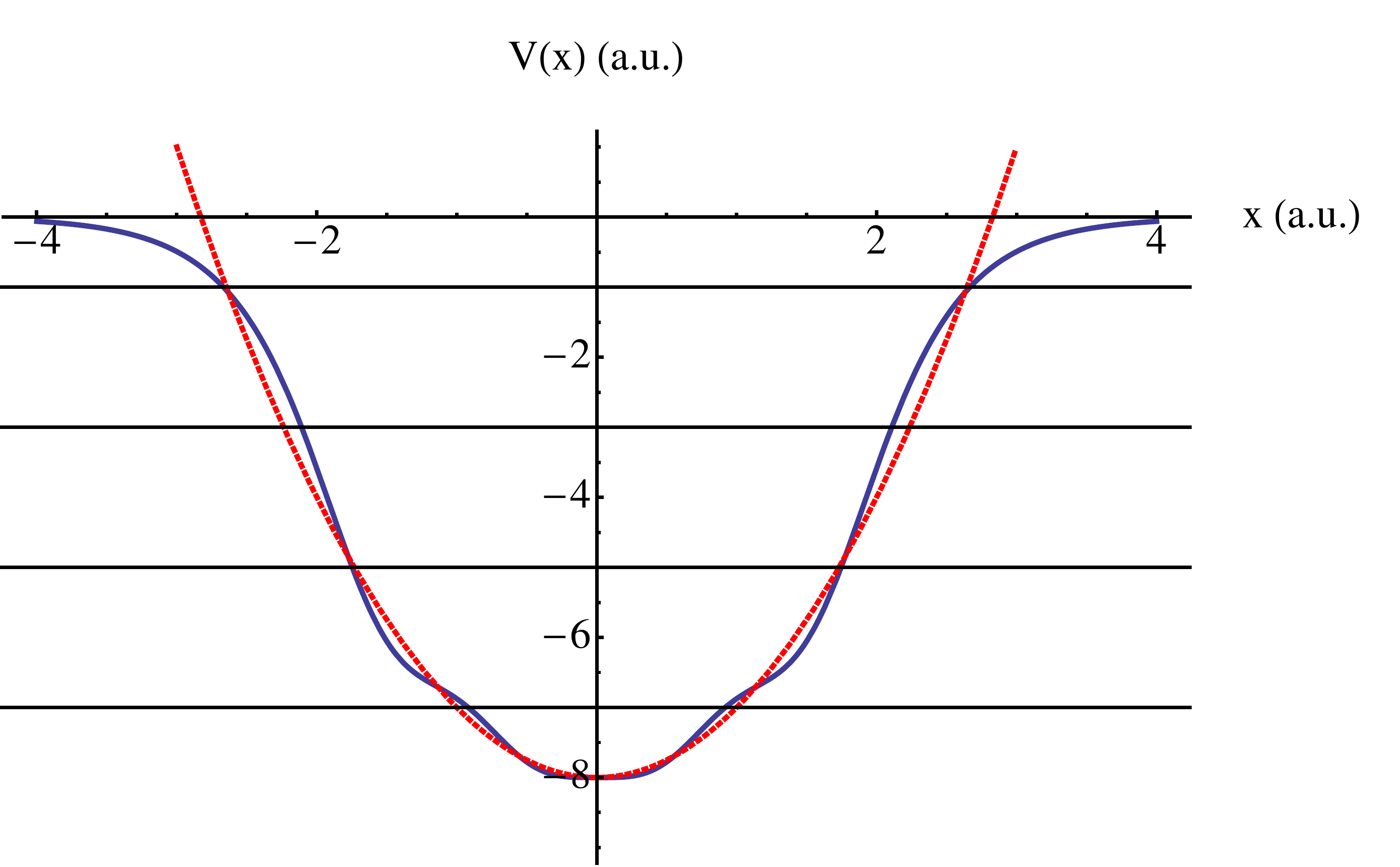}}
\caption{\label{shoV6} The solid curve is the reflectionless potential with only four energy levels that match the first four of a quantum simple harmonic oscillator, scaling like $n$ relative to the ground state.  The horizontal lines are the associated energy levels of the solid curve potential.  The dotted curve is the exact simple harmonic oscillator potential with a spring constant $k=2$.  } 
\end{figure}

\begin{figure}[h!]
\resizebox{.5\textwidth}{!}{\includegraphics{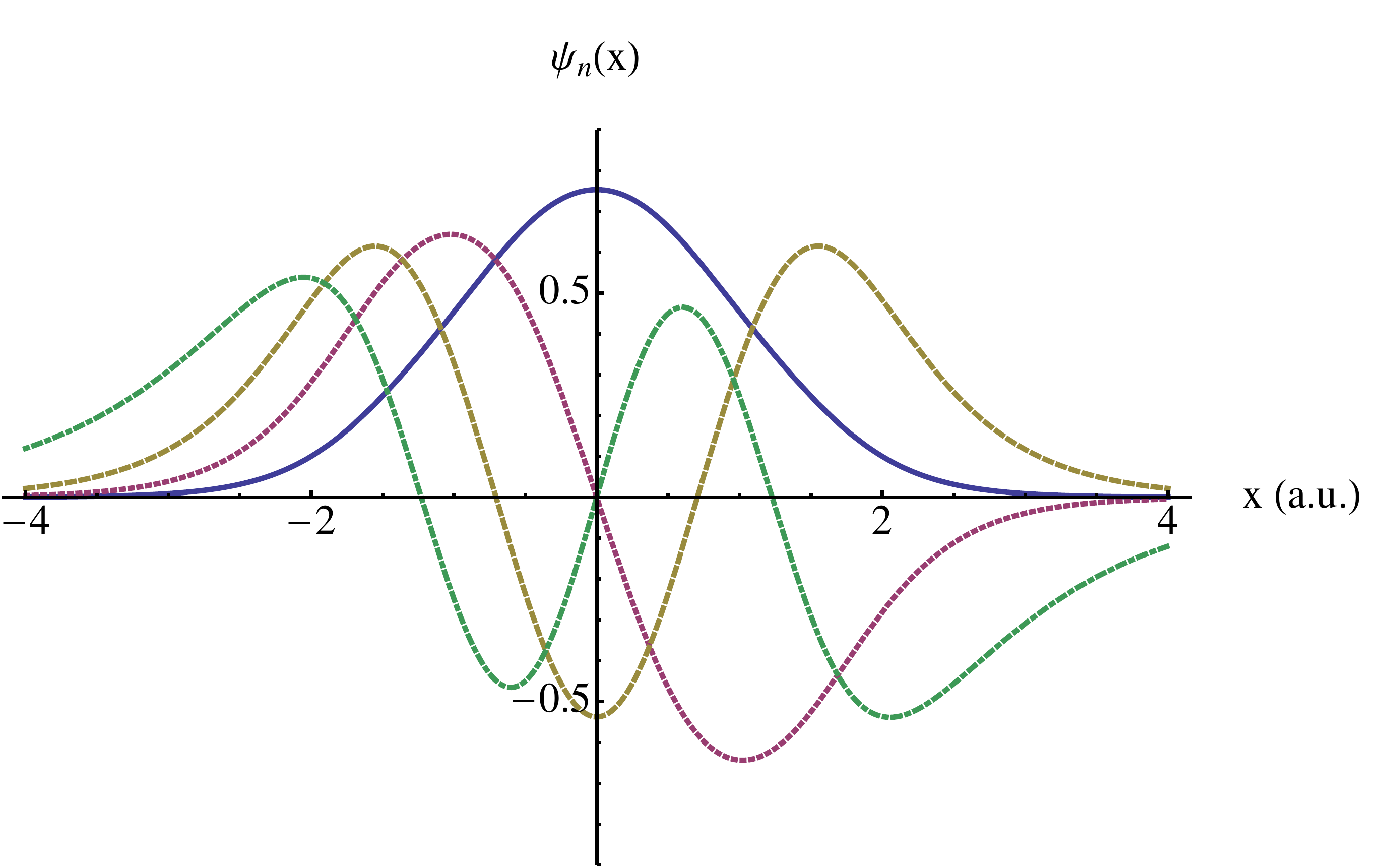}}
\caption{\label{shoWF3} The first four normalized bound state wave functions of the potential in Fig.~\ref{shoV6} generated via the Kay and Moses method described in the text.  The solid curve is the ground state, the dotted curve the first excited state, the dashed curve the second excited state, and the dash-dot curve the third excited state. This should be compared to the corresponding exact wave functions for the simple harmonic oscillator in Fig.~\ref{shoWFX}. } 
\end{figure}

\begin{figure}[h!]
\resizebox{.5\textwidth}{!}{\includegraphics{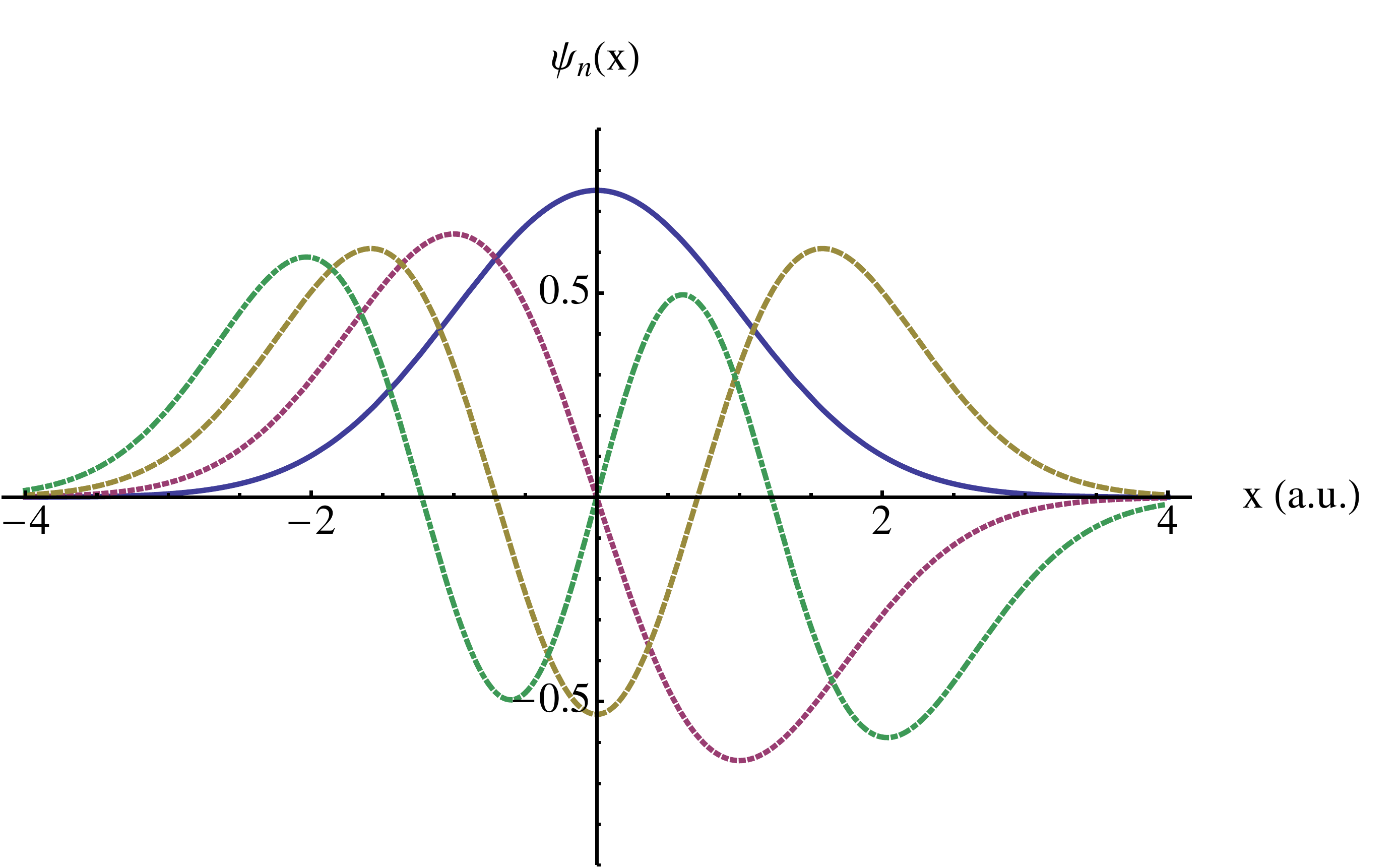}}
\caption{\label{shoWFX} The first four normalized bound state wave functions of the simple harmonic oscillator with $k=2$ using the same parameters in Fig.~\ref{shoWF3}.  The solid curve is the ground state, the dotted curve the first excited state, the dashed curve the second excited state, and the dash-dot curve the third excited state. } 
\end{figure}

The simple harmonic oscillator has evenly spaced levels relative to the ground state.   Figure \ref{shoV6} shows the reflectionless potential generated using the Mar\u{c}henko method.  Its four bound states are constructed to match the first four energy levels of the simple harmonic oscillator with spring constant $k=2$ in some units.  Figure \ref{shoWF3} shows the first four normalized wave functions generated using the Kay and Moses method discussed above and Fig.~\ref{shoWFX} shows the first four exact wave functions for the equivalent simple harmonic oscillator potential (the dotted curve in Fig.~\ref{shoV6}).  The shapes of the wave functions closely match those of the exact solutions.  By inspection, the shape of the potential appears as only a perturbation on the actual simple harmonic oscillator system, so this is not entirely unexpected.  However, because the reflectionless potential in Fig.~\ref{shoV6} is a finite potential, the tails of those solutions extend further than the equivalent states of the exact solutions.  The effect becoming exaggerated for excited states, which are more sensitive to the edge of the potential.

\begin{figure}[h!]
\resizebox{.5\textwidth}{!}{\includegraphics{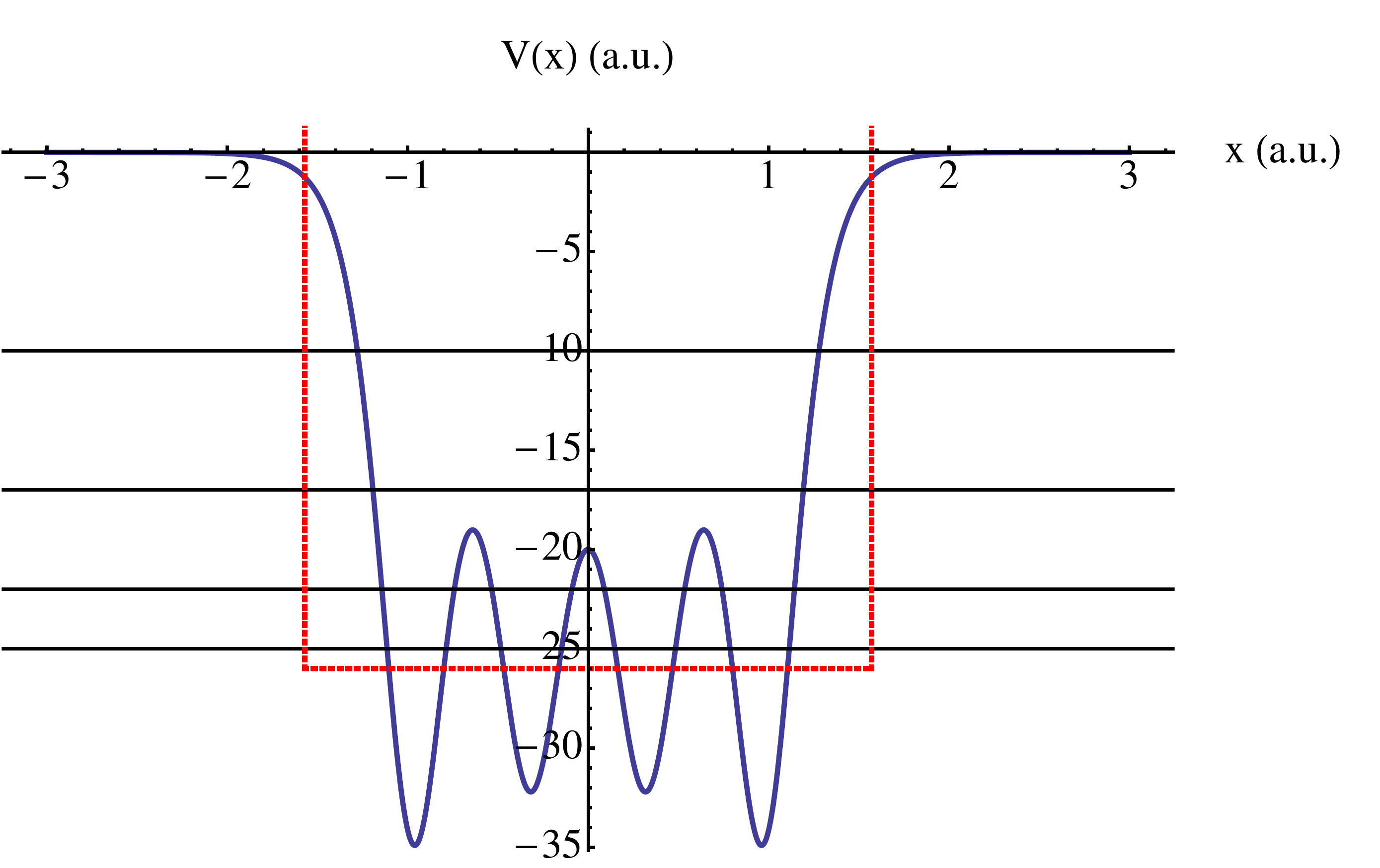}}
\caption{\label{isw}The solid curve is the reflectionless potential with only four energy levels that match the first four of an infinite square well, scaling like $n^2$ relative to the ground state.  The horizontal lines are the associated energy levels of the solid curve potential.  The dotted lines represent the equivalent infinite square well whose walls are at $\pm\pi/2$ with a base potential of $V_0=-26$ in some units.} 
\end{figure}

\begin{figure}[h!]
\resizebox{.5\textwidth}{!}{\includegraphics{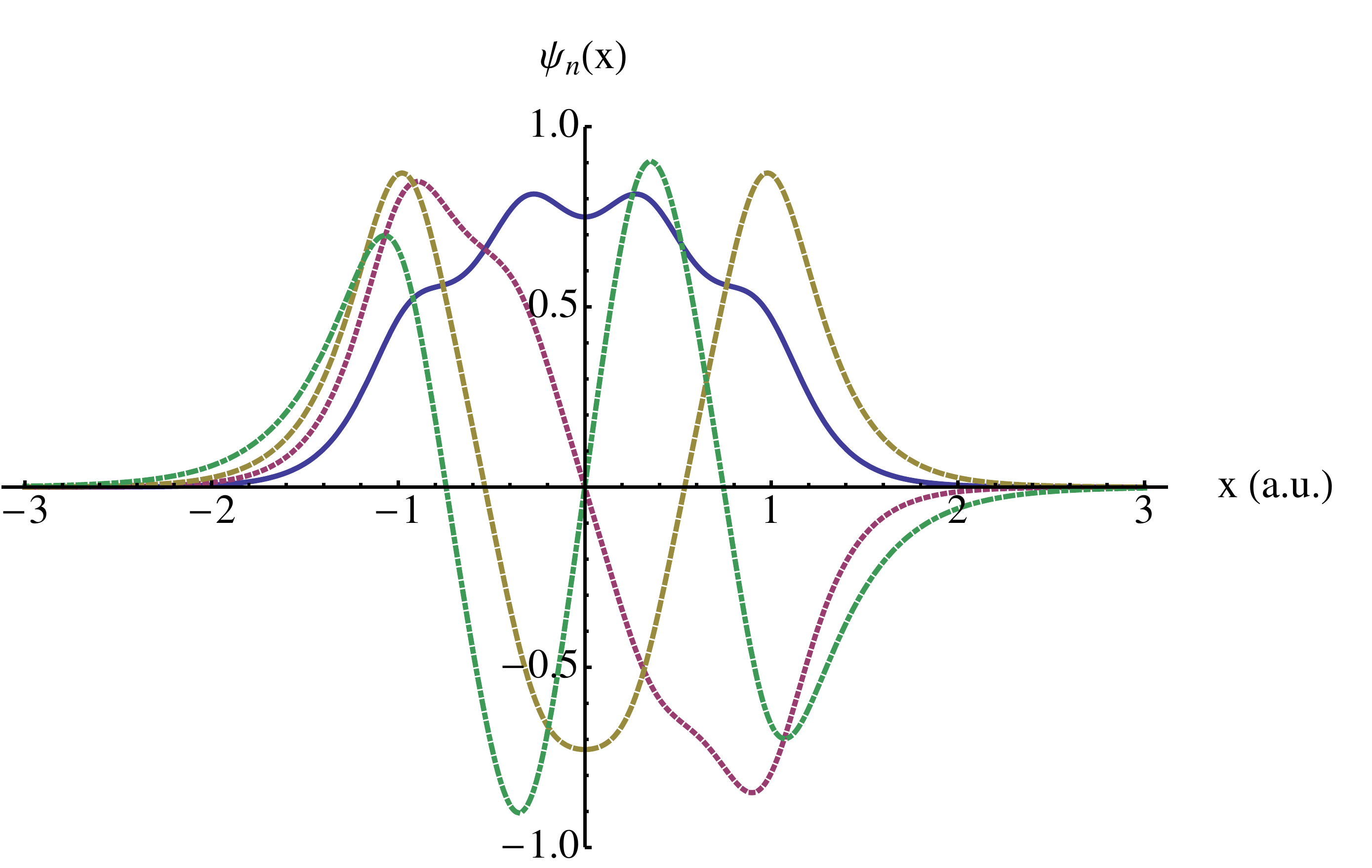}}
\caption{\label{iswwf} The first four normalized bound state wave functions of the potential in Fig.~\ref{isw}, whose energies match the first four states of the infinite square well centered at the origin whose width is $L=\pi$ and base potential $V_0=-26$ in appropriate units.  These were generated via the Kay and Moses method described in the text.  Notice that the low lying states only crudely resemble the the infinite square well wave functions because they can ``feel'' the lumpy bottom approximating the potential's shape.  But as the states become more excited, they begin to resemble the expected wave functions. The solid curve is the ground state, the dotted curve the first excited state, the dashed curve the second excited state, and the dash-dot curve the third excited state.  This should be compared to the corresponding exact wave functions for the infinite square well in Fig.~\ref{iswwfx}. } 
\end{figure}

\begin{figure}[h!]
\resizebox{.5\textwidth}{!}{\includegraphics{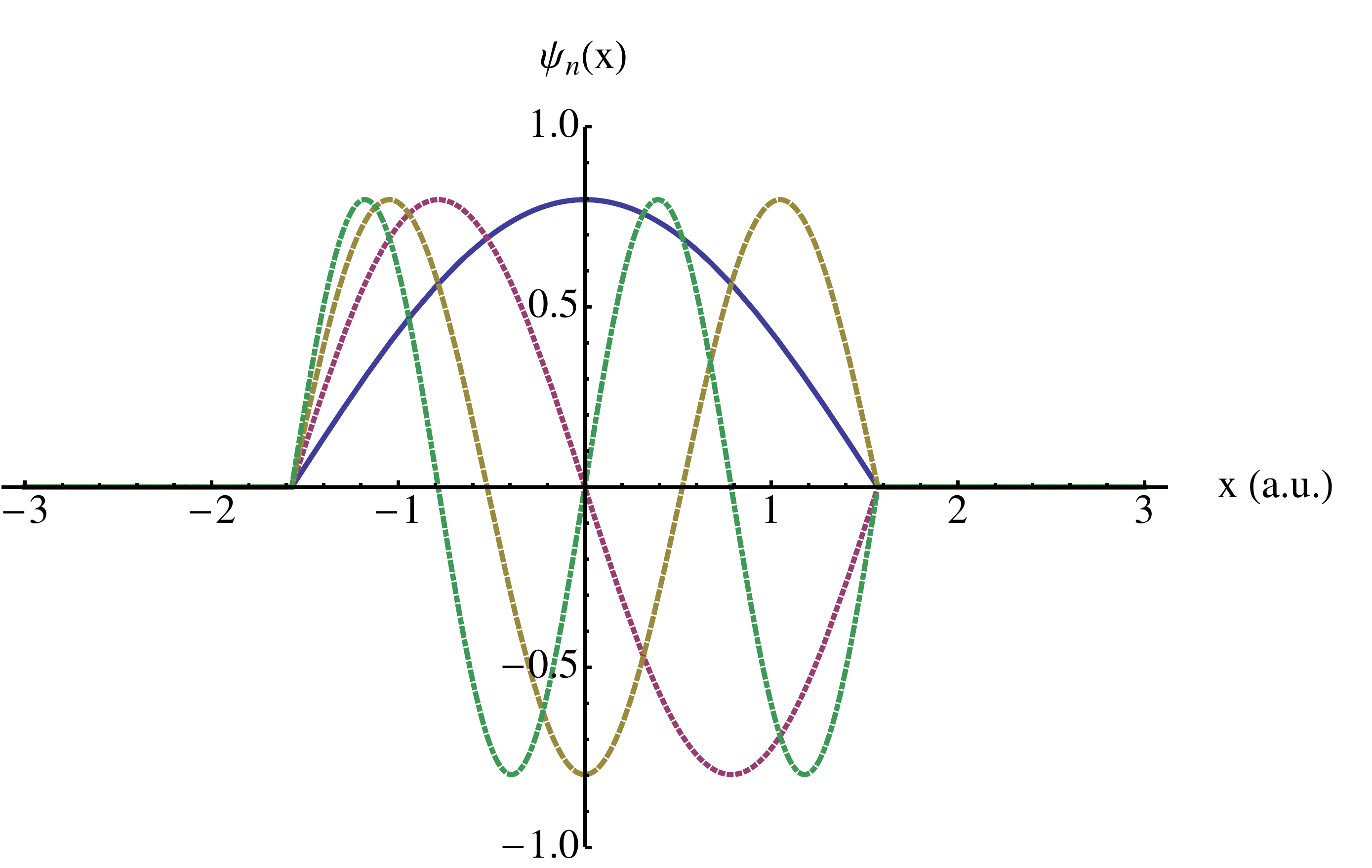}}
\caption{\label{iswwfx} The first four normalized bound state wave functions of the infinite square well centered at the origin whose width is $L=\pi$ in appropriate units.  The solid curve is the ground state, the dotted curve the first excited state, the dashed curve the second excited state, and the dash-dot curve the third excited state.  This should be compared to the corresponding states in Fig.~\ref{iswwf}. } 
\end{figure}

The infinite square well has energy levels that scale like $n^2$ with respect to the ground state.  Figure \ref{isw} shows the reflectionless potential generated using the  Mar\u{c}henko method.  The four bound state energies match the first four energies of an infinite square well centered at the origin whose width is $L=\pi$ in appropriate units.  Figure \ref{iswwf} shows the first four normalized wave functions.  Figure \ref{iswwfx} shows the exact solutions.  In contrast to the simple harmonic oscillator, this potential only superficially represents the original system, but is a plausible analog considering the potential here is finite.  The rapid oscillations at the bottom of the well serve as a ``flat bottom'' that conspire with the tapered boundaries to generate the appropriate features of the energy spectra.  The associated wave functions suffer distortions compared to the exact solutions; they are sensitive to the lumpy structure and their tails extend beyond the confines of the infinite square well.

\begin{figure}[h!]
\resizebox{.5\textwidth}{!}{\includegraphics{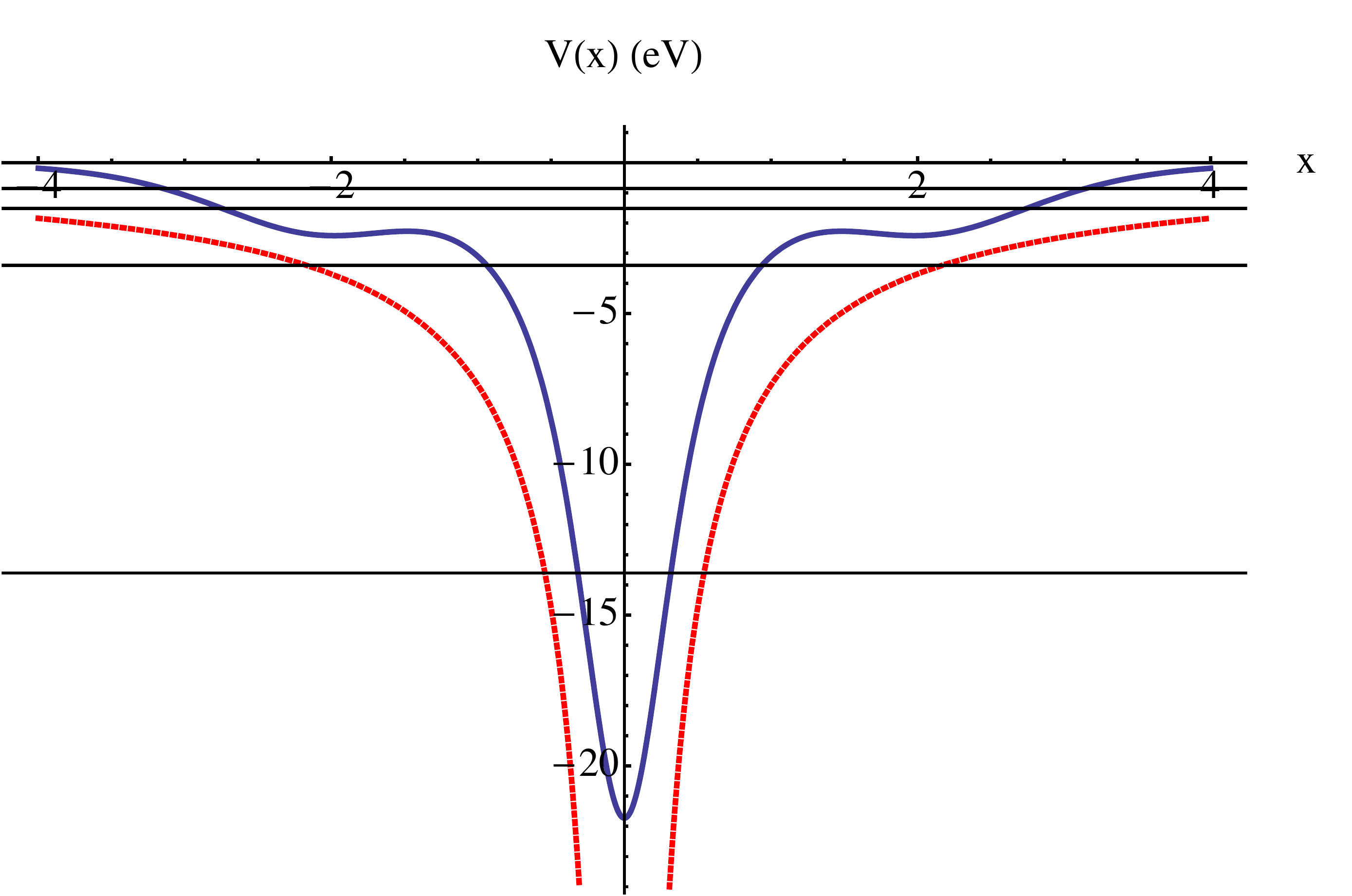}}
\caption{\label{h16} The solid curve is the reflectionless potential with four energy levels that match the first four finite energies for the one-dimensional hydrogen atom.  They scale like $1/n^2$ relative to the ground state.  The horizontal lines are the associated energy levels of the solid curve potential.  The energy units are in eV and each distance unit along the $x-$axis is about 2~\r{A} (see Exercise 1 below). The dotted curve is the exact potential for the one-dimensional hydrogen atom. } 
\end{figure}

\begin{figure}[h!]
\resizebox{.5\textwidth}{!}{\includegraphics{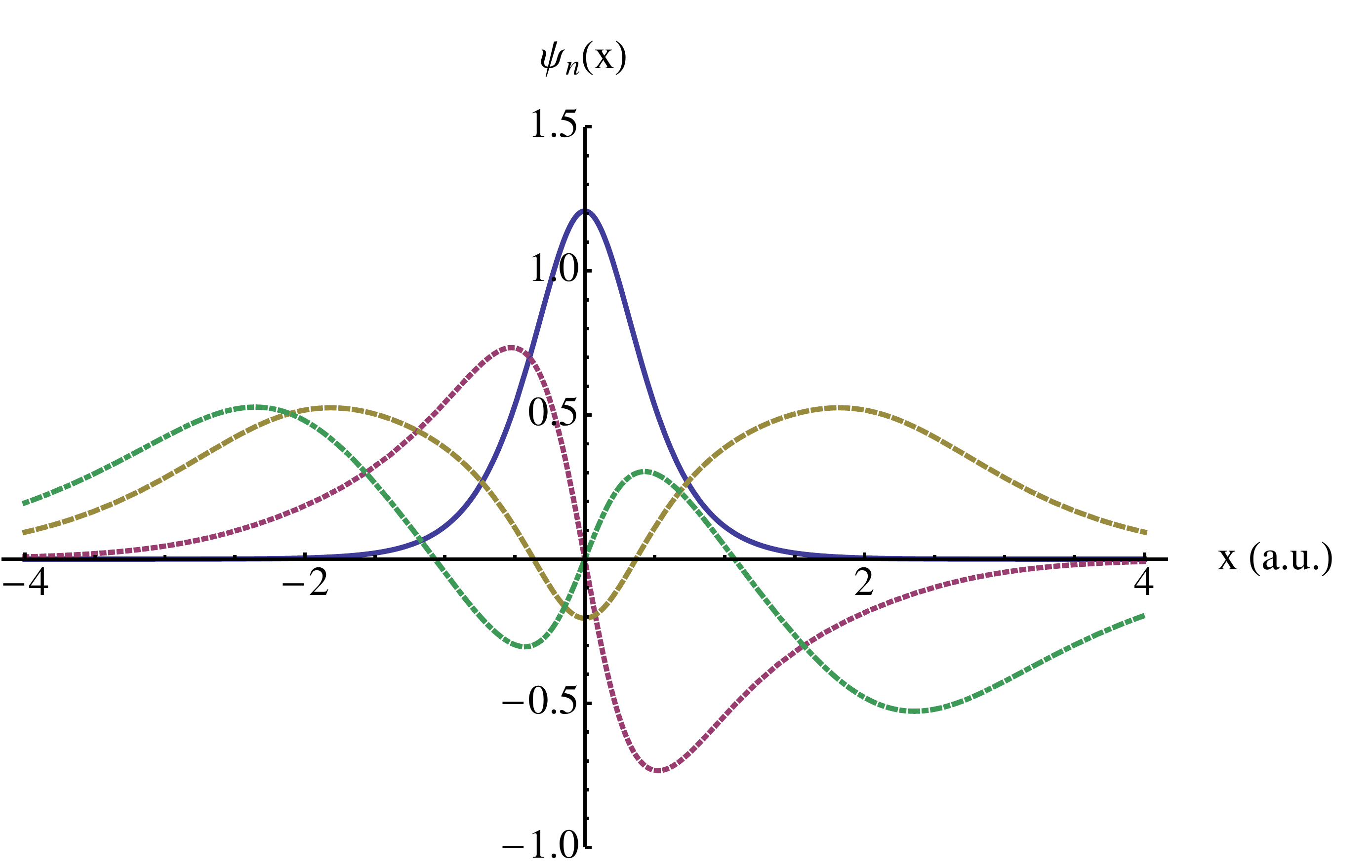}}
\caption{\label{h1wf} The first four normalized bound state wave functions of the potential in Fig.~\ref{h16}  generated via the Kay and Moses method described in the text.  The solid curve is the ground state, the dotted curve the first excited state, the dashed curve the second excited state, and the dash-dot curve the third excited state. } 
\end{figure}

\begin{figure}[h!]
\resizebox{.5\textwidth}{!}{\includegraphics{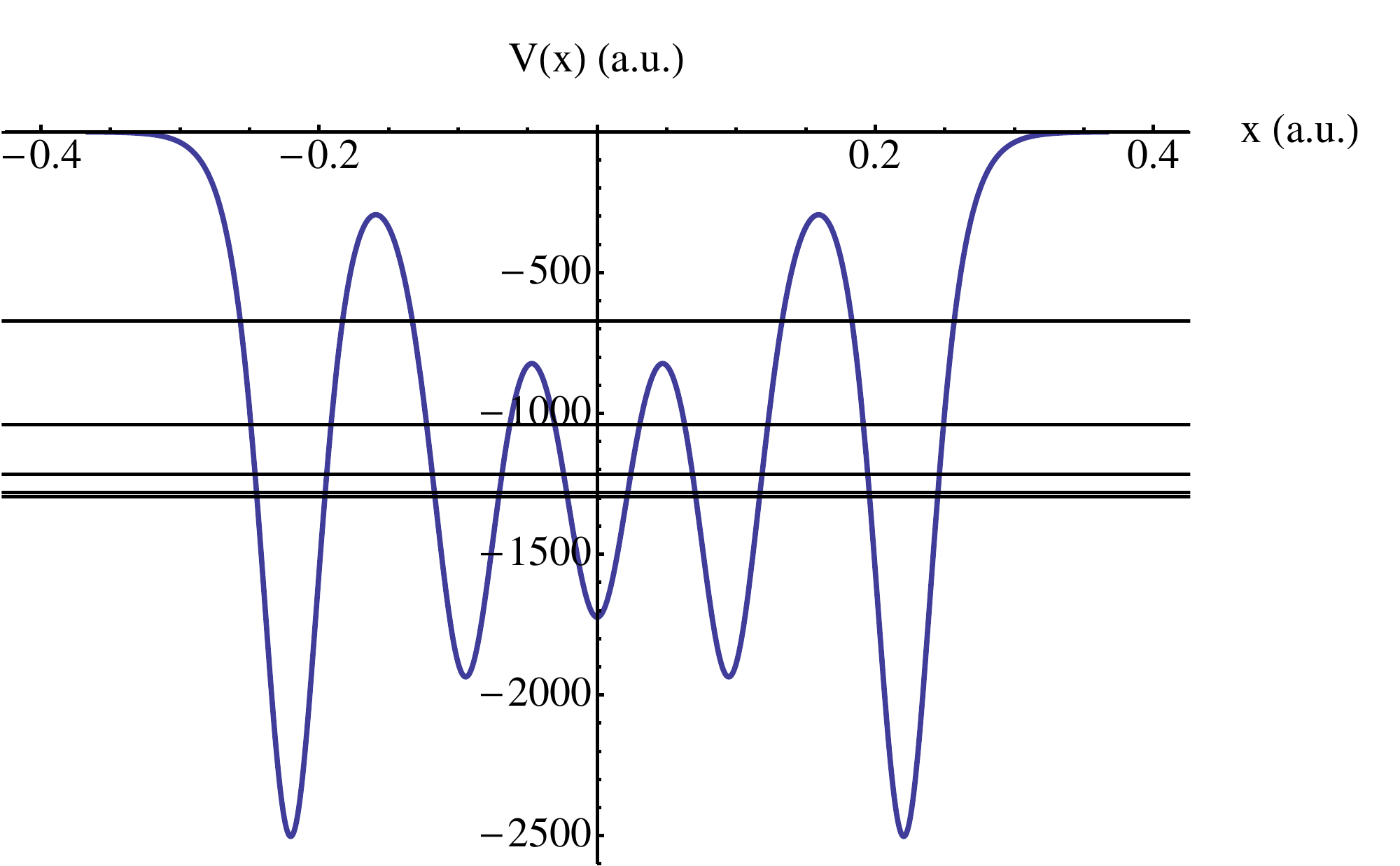}}
\caption{\label{p4} A reflectionless potential having five energy levels that scale like $n^4$ relative to the ground state. The horizontal lines are the associated energy levels of the solid curve potential.  The two lowest energy levels are very closely spaced on this scale. } 
\end{figure}

\begin{figure}[h!]
\resizebox{.5\textwidth}{!}{\includegraphics{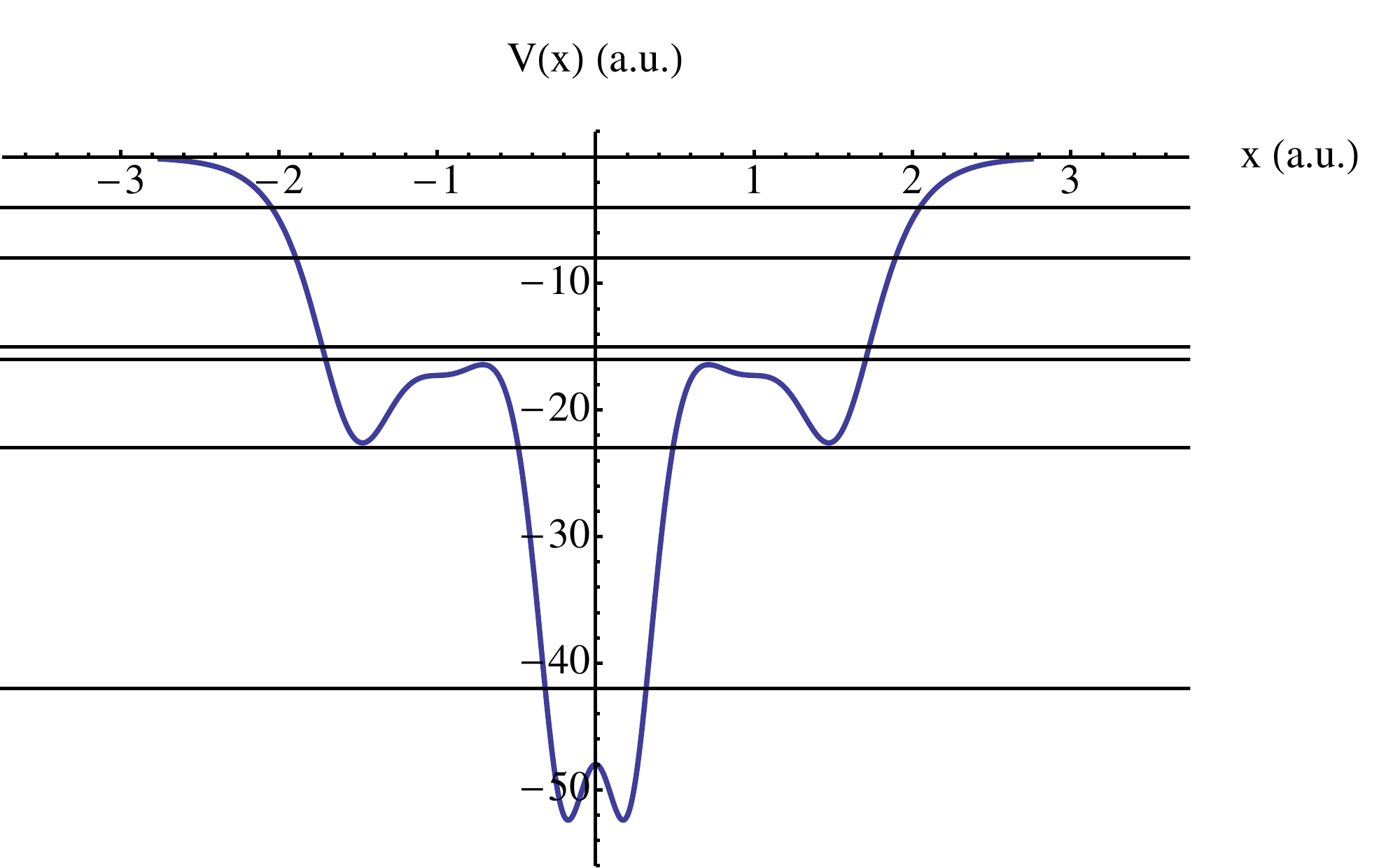}}
\caption{\label{lost1} A reflectionless potential having energy eigenvalues that correspond to negative values of the six numbers from the television show {\em Lost}: -4, -8, -15, -16, -23, and -42.  The horizontal lines are the associated energy levels of the solid curve potential.  } 
\end{figure}

The one-dimensional hydrogen atom is an interesting system, analytically explored in detail by Loudon \cite{Loudon55}.  Although straightforward, some care must be taken in interpreting the results.   The one-dimensional hydrogen atom has a set of finite energy levels that scale like $1/n^2$ with respect to the lowest finite-energy state, generating the familiar Rydberg energies.  However, curiously, the ground state has an even wave function very tightly localized at the origin but has infinite binding energy.   Otherwise, the one-dimensional system shares the bound state energy spectra for three-dimensional hydrogen.  That is, its ``first excited state'' has a finite energy that matches the familiar ground state energy of the three-dimensional case.  Moreover, all of the finite energy states are doubly degenerate with both and even and odd wave functions about the origin.  This seems to violate the non-degeneracy theorem, which states that one-dimensional systems cannot have degenerate states.  However, as is shown in \cite{Loudon55}, this only pertains to potentials which have no singular points and so does not apply to one-dimensional hydrogen potential with a singularity at the origin.

What is produced by the methods here more resembles the wave functions for Loudon's truncated system, which regulates the $1/|x|$ behavior at some cutoff scale $x_0$.  Figure \ref{h16} shows the reflectionless potential that matches the first four finite energies of the one-dimensional hydrogen atom, which are the same as the first four familiar hydrogen atom Rydberg energies.  The first four normalized bound state wave functions of the potential in Fig.~\ref{h16} are shown in Fig.~\ref{h1wf}.  Consistent with the non-degeneracy theorem, the reflectionless hydrogen atom has no singularities, so has no degeneracies.  The value of the ground state was chosen to be -13.6 eV.  However, recall that the units being used are such that $\hbar^2/(2m)=1$, so the position units must be interpreted carefully (see Exercise 1 below).

Note now, in particular, the ground state wave function in Fig.~\ref{h1wf}.  Its derivative at the origin is zero, consistent with a tacitly enforced smoothness criteria across the origin of the $x$-axis for this non-singular potential.   This reflectionless hydrogen atom ground state has a finite energy and a probability per unit length that peaks at the origin.  This is in contrast to the actual one-dimensional hydrogen atom, which has infinite binding energy.   This is also unlike the three-dimensional ground state, which has a probability per unit volume that peaks at the origin, but a radial probability that peaks at the Bohr radius.  The ground state for the reflectionless one-dimensional case has a genuine maximum probability per unit length at the origin and does not peak at an equivalent Bohr radius.

It is perhaps worth pausing to admire the reflectionless hydrogen atom system.  Here is a one-dimensional potential whose bound states generate the usual set of low lying spectral series for hydrogen, yet is completely transparent to matter waves of all energies impinging on it.

An interesting case to consider is when the energy level scaling goes like $n^p$ where $p>2$.   For an infinite number of states in a confined potential, it can be shown that the energy levels cannot scale faster than $n^2$, i.e. at a rate faster than that of an infinite square well.  This result can be demonstrated using WKB methods  \cite{Muss97,Schum08}.  The fundamentals of the powerful WKB method are covered in many undergraduate courses on quantum mechanics \cite{Griffiths05}.  The result has intuitive value: regardless of your potential, you cannot squeeze the quantum energy eigenstate distribution faster than the states generated by the infinite square well, which is, in this context, maximally confining.  This limitation is true for bounding potentials that contain an infinite number of energy states.  For example, as indicated above, there is no infinitely bound $V(x)$ in one dimension that can have an energy spectrum scaling like $n^3$.    However, the Mar\u{c}henko method can be applied for a finite set of spectra that scale faster than $n^2$ relative to the ground state.  Figure \ref{p4} shows the shape of the potential having five energy levels that scale like $n^4$.

In the exercises in Sec.~\ref{ex}, students are encouraged to explore the forms for reflectionless potentials and associated wave functions for energy spectra of their own choosing.  As an example of a custom spectrum, Fig.~\ref{lost1} shows the potential having six energy eigenvalues corresponding to the recurring mysterious numbers from the TV show 
\textit{Lost}~\cite{losttv}:  \{4, 8, 15, 16, 23, 42\}.  Note, in keeping with the procedure, the values of the numbers as energies are taken as negative so -42 is the ground state.

\subsection{Conclusion}
The inverse bound state problem is a fascinating formal problem in quantum theory.  While the nuanced mathematics can be rather involved, the core mathematical technology is accessible to the motived undergraduate.  Here, an algorithm for implementing the Mar\u{c}henko method was outlined that generates an analytical form for a reflectionless potential from a finite set of energy eigenstates.  Another algorithm was discussed, introduced by Kay and Moses, that produces the associated analytical scattering and bound state wave functions.   Hopefully, instructors and students alike will find pedagogical value in experimenting with these methods.

\appendix
\section{Exercises}\label{ex}
\begin{enumerate}
%\subsubsection*{Exercise 1}
\item Throughout this paper, units where $\hbar^2/2m=1$ were used.  Convenient natural unit choices of this kind are common in theoretical physics, but can be confusing on first exposure, particularly when converting to more familiar unit systems.  Show that $L=\pi$ in some natural distance unit for the infinite square well problem investigated in this paper. Like in Fig.~\ref{isw}, use $V_0=-26$ and $E_1=1$ (with respect to the bottom of the well) expressed in some natural energy unit.  Without more information, is it possible to express $L$ in meters, for example? Express $L$ in femtometers if it is known the particle in the infinite well is a pion with mass 140 MeV/$c^2$? For the hydrogen atom problem, express the implied distance units discussed in the text in nanometers given that we are working with energy in eV for an electron in a Coulomb potential. 

\item Work through the procedure as outlined in Sec.~\ref{sum} for the case with $N=1$; keep $\kappa_1$ as an unspecified variable.  This represents the simplest reflectionless potential with one bound state. Explore the functional form of the potential, $V(x)$, the bound state wave function, $\psi_1(x)$,  and the scattering state wave function, $\Psi(x,E)$.  Arrange your exponentials in terms of convenient hyperbolic trig functions and simplify.  Show that the transmission coefficient is unity.  Compare $\Psi(x,E)$ and $V(x)$ to the reflectionless potential and wave function from Problem 2.51 in {\em Introduction to Quantum Mechanics} by Griffiths \cite{Griffiths05}.  Do the exercise by hand, but check your work with a computer.

%\subsubsection*{Exercise 2}
\item Using your favorite mathematical software, follow the procedure as outlined in Sec.~\ref{sum}  to reproduce the examples in this paper.  How many states are you able to find analytical solutions for before the solutions become unstable or the computation takes more than a few minutes (results will vary depending on your computer)?

%\subsubsection*{Exercise 3}
\item Using your favorite mathematical software, follow the procedure as outlined in Sec.~\ref{sum} to produce the associated $V(x)$ and $\psi_n(x)$ for your own set of $N$  $\kappa_n$.  Like the graphs in this paper, plot the wave function, potential, and the energy levels and comment on any interesting features or structures.  Do the solutions seem plausible? By hand, but checking your result with a computer, verify that your ground state wave function, $\psi_1(x)$, satisfies Schr\"{o}dinger equation for the bound states (Eq.~\ref{se}) consistent with your number of states, $N$, your potential, $V(x)$, and your selected $\kappa_1$.  Sample suggestions for energy eigenvalues: the first few Fibonacci numbers sans the first 1 with $-E_n\sim\{1, 2, 3, 5, 8, 13\}$; the first few notes of the equal tempered music scale with $-E_n\sim 2^{(n/12)}$.  Note the minus sign on $E_n$.  As mentioned in the text, obtaining the potentials for six or seven states should be straightforward, but extracting the wave functions might be numerically unstable beyond four states.

%\subsubsection*{Exercise 4}
\item Plug the solution for $\Psi(x,E)$ (Eq.~\ref{scatwf}) into the Schr\"{o}dinger equation for the scattering states (Eq.~\ref{scat}) and show that $\psi_n(x)$ satisfies Schr\"{o}dinger equation for the bound states (Eq.~\ref{se}) if $E_n=-\kappa_n^2$ and the form of the potential is $V(x)=2\frac{d}{dx}\sum_{n=1}^N\psi_n(x)e^{\kappa_n x}$. Do the exercise by hand, but check your work with a computer.\label{deriv}

%\subsubsection*{Exercise 5}
\item Assuming $\lim_{x\rightarrow\infty}\psi_n(x)\sim c_n^2 e^{-\kappa_n x}$,  investigate the asymptotic form of Eq.~\ref{km} and derive Eq.~\ref{cn} for the normalization constants $c_n^2$.  Do the exercise by hand, but check your work with a computer.
\end{enumerate}

\section{Solutions and hints for selected exercises}\label{sol}
\subsubsection*{Solution to Exercise 1}
Although language like ``units of $\hbar^2/2m$'' is commonly used, it should be appreciated that the choice of $\hbar^2/2m=1$ fixes the unit relationships but does not by itself fix a particular unit system. More information is required to complete the conversion.  In the case of the infinite square well, the energy spectra are given by 
\begin{equation}
E_n=\frac{\hbar^2\pi^2}{2mL^2}n^2.  \label{iswE}
\end{equation}  
In the problem as presented, the ground state energy, $E_1$, has ``one energy unit'' with respect to the bottom of the well (look at Fig.~\ref{isw}).  These energy units are not themselves specified.  However, if $E_1=1$ and $\hbar^2/2m=1$ then, substituting these into Eq.~\ref{iswE}, $L=\pi$ in a still unspecified length unit, which will be called ``q'' below.  This is most that can be said.  However, if the mass of particle in the well was known, for example, this would then fix an energy scale for the problem and any explicit system of units can be assigned if desired.  In this problem, the pion has a rest energy of $m_\pi c^2=140$~MeV, so it is convenient to select MeV as an energy unit.  It is helpful to write $\hbar^2/2m=1$ by multiplying the top and bottom by $c^2$.  This gives,
\begin{equation}
(\hbar c)^2=2mc^2
\label{hb}
\end{equation}
or
\begin{equation}
\hbar c=\sqrt{(2)(140)}~{\rm MeV}\cdot{\rm q}=16.7~{\rm MeV}\cdot{\rm q}.\label{hb2}
\end{equation}
Notice that the units of $(\hbar c)^2$ in Eq.~\ref{hb} appear to carry units of energy explicitly.  However, in Eq.~\ref{hb2}, the units are energy$\cdot$distance as always.  Equation \ref{hb} indicates to configure the units of energy and distance such that the numerical values of $(\hbar c)^2$ are the same as $2mc^2$.  This means if the energy units are selected explicitly, then the distance units must shift to accommodate.  A conversion of q units to femtometers can be done by noting that $\hbar c=197$~MeV$\cdot$fm, so q length units are about 11.8 fm.
The width of the well is then 
\begin{equation}
L=\pi~{\rm q}=(\pi)(11.8)~{\rm fm}=~37.1~{\rm fm}.
\end{equation}
In the hydrogen atom problem the energy units have been fixed as eV, as can be seen in Fig.~\ref{h16} by the choice of -13.6 eV for the ground state energy.  To determine the length units in nanometers, previous knowledge of the hydrogen atom parameters in known units are used.	Now, let ``r'' represent the name for the unknown length unit. The ground state energy can be written in terms of fundamental parameters as
\begin{equation}
E_R=13.6~{\rm eV}=\frac{m_e c^2}{2 (\hbar c)^2}(ke^2)^2
\label{er}
\end{equation}
with $k=1/4\pi\epsilon_0$.  Since $\hbar^2/2m=1$, Eq.~\ref{er} can be rearranged and solved for $ke^2$
\begin{equation}
ke^2=\sqrt{4 E_R}=\sqrt{(4)(13.6)}~{\rm MeV}\cdot{r}=7.38~{\rm eV}\cdot{r}.\label{ke}
\end{equation}
At a glance, Eq.~\ref{ke} seems to imply that $ke^2$ now carries units of $\sqrt{{\rm energy}}$.  However, this equation indicates a numerical relationship and there are hidden units;  $ke^2$ still has units of energy$\cdot$distance, in this case MeV$\cdot$r.  Now, knowing that $ke^2=1.44$~eV$\cdot$nm, Eq.~\ref{ke} can be used to determine that r length units are about 0.195 nm.
         
\subsubsection*{Solution to Exercise 2}
First, select one energy eigenvalue $E_1=-\kappa_1^2$ where the constant $\kappa_1>0$.  There will be one normalization constant, using Eq.~\ref{cn}, 
\begin{equation}
c_1^2=2\kappa_1.
\label{cex}
\end{equation}
There will be one matrix element for the one-element ``matrix'' ${\bf C}$ given by Eq.~\ref{CC}
\begin{equation}
C_{11}=\frac{c_1^2}{2\kappa_1}e^{-2\kappa_1 x}=e^{-2\kappa_1 x}.
\end{equation}  
To construct $V(x)$ from Eq.~\ref{mV} it will be helpful to first construct the matrix 
\begin{equation}
{\bf I}+{\bf C}=\det{({\bf I}+{\bf C})}=1+e^{-2\kappa_1 x}
\end{equation} 
(being equal to its own determinant since it is a one-element ``matrix'').  Next, it will be helpful to construct the derivative
\begin{equation}
\frac{d}{dx}\ln{\det{({\bf I}+{\bf C})}}=\frac{-2\kappa_1 e^{\kappa_1 x}}{1+e^{-2\kappa_1 x}},
\label{d1}
\end{equation}
recalling that
\begin{equation}
\frac{d}{dx}\ln{f(x)}=\frac{1}{f(x)}\frac{df(x)}{dx}
\end{equation}
with $f(x)=\det{({\bf I}+{\bf C})}$.

Finally, taking another derivative of Eq.~\ref{d1} gives
\begin{equation}
\frac{d^2}{dx^2}\ln{\det{({\bf I}+{\bf C})}}=4\kappa_1^2\left[
\frac{e^{-4\kappa_1 x}}{(1+e^{-2\kappa_1 x})^2}-\frac{e^{-2\kappa_1 x}}{1+e^{-2\kappa_1 x}}
\right]
\end{equation}
so, plugging into Eq.~\ref{mV},
\begin{equation}
V(x)=-8\kappa_1^2\left[
\frac{e^{-4\kappa_1 x}}{(1+e^{-2\kappa_1 x})^2}-\frac{e^{-2\kappa_1 x}}{1+e^{-2\kappa_1 x}}
\right].
\end{equation}
Simplifying, using 
\begin{equation}
{\rm sech}(\kappa_1 x)=\frac{2}{e^{\kappa_1 x}+e^{-\kappa_1 x}},
\end{equation}
gives
\begin{equation}
V(x)=-2\kappa_1^2~{\rm sech}^2(\kappa_1 x),
\label{v1}
\end{equation}
which matches the form of the reflectionless potential in Griffiths problem 2.51 (accounting for the difference in units; recall here that $\hbar^2/2m=1$).   This is a special case of what is sometimes called a modified P\"{o}schl-Teller or ``Sech-squared'' potential \cite{Kiri98}.

To obtain $\psi_1(x)$, plug $c_1^2$ from Eq.~\ref{cex} into Eq.~\ref{km} and solve for $\psi_1(x)$ giving
\begin{equation}
\psi_1(x)=-\frac{-2\kappa_1 e^{\kappa_1 x}}{1+e^{2\kappa_1 x}}=-\kappa_1~{\rm sech}(\kappa_1 x),
\label{wfex}
\end{equation}
with the normalized bound state wave function being
\begin{equation}
\frac{\psi_1(x)}{c_1}=-\sqrt{\frac{\kappa_1}{2}}~{\rm sech}(\kappa_1 x).
\end{equation}
The minus sign, an overall phase factor, does not affect the physical picture.  Note that Eq.~\ref{km} here only has one equation and one unknown, $\psi_1(x)$, and the sum over $\nu$ is only one term.  

To obtain the scattering wave function, $\Psi(x,E)$, plug the result from Eq.~\ref{wfex} into Eq.~\ref{scatwf}
\begin{equation}
\Psi(x,E)=\left[1-\frac{\kappa_1~{\rm sech}(\kappa_1 x)}{\kappa_1+i\sqrt{E}}e^{\kappa_1 x}\right]e^{i\sqrt{E}x}.
\end{equation}
The equation can be simplified using 
\begin{equation}
{\rm tanh}(\kappa_1 x)=\frac{e^{\kappa_1 x}-e^{-\kappa_1 x}}{e^{\kappa_1 x}+e^{-\kappa_1 x}},
\end{equation}
and letting $k=\sqrt{E}$.  After some algebra,
\begin{equation}
\Psi(x,E)=\left[\frac{ik-\kappa_1\tanh{(\kappa_1 x)}}{\kappa_1+ik}\right]e^{ikx},
\label{scatex}
\end{equation}
which matches the form of the scattering wave function in problem 2.51 from Griffiths.  The transmission coefficient is obtained by looking at the asymptotic behavior of $|\Psi(x,E)|^2$ as $x\rightarrow\infty$ (assuming an incident beam from the left). In this limit, Eq.~\ref{scatex} behaves as
\begin{equation}
\lim_{x \to +\infty}\Psi(x,E)=\left[\frac{ik-\kappa_1}{\kappa_1+ik}\right]e^{ikx},
\label{lim1}
\end{equation}
so the transmission coefficient is
\begin{equation}
\lim_{x \to +\infty}|\Psi(x,E)|^2=1.
\label{lim2}
\end{equation}
This implies (by unitarity) that the reflection coefficient is 0, which can be seen by inspection since there is no $e^{-ikx}$ term in Eq.~\ref{scatex}.  But more to the point, reflectionlessness was a feature by design because of the form of Eq.~\ref{scatwf}.  This problem is treated in \cite{Kay56}, but be mindful of the typographical error in the form of the potential in their equation 4.8 where it is written $V(x)=-2\kappa_1^2~{\rm sech}^2(2\kappa_1 x)$ rather than as Eq.~\ref{v1} above.
%\subsubsection*{Exercise 2}
%
%\subsubsection*{Exercise 3}

\subsubsection*{Hints to Exercise 5}
This is a fairly tedious exercise, even though each mathematical step is fairly straightforward.  After plugging $\Psi(x,E)$ (Eq.~\ref{scatwf}) into the Schr\"{o}dinger equation for the scattering states (Eq.~\ref{scat}) and, after some manipulation, the result can be put into the form
\begin{equation}
\left\{\left[\sum_{n=1}^N\frac{(\hat{M}_n\psi_n) e^{\kappa_n x}}{\kappa_n+i\sqrt{E}}\right]+\left[2\frac{d}{dx}\sum_{n=1}^N\psi_n e^{\kappa_n x}-V\right]\right\}e^{i\sqrt{E}x}=0
\label{ex4}
\end{equation}
where the operator 
\begin{equation}
\hat{M}_n=\frac{d^2}{dx^2}-\kappa_n^2-V(x)
\end{equation}
is defined so that the Schr\"{o}dinger equation for the bound states, Eq.~\ref{se}, can be expressed as
\begin{equation}
\hat{M}_n\psi_n=0.
\label{se2}
\end{equation}
Notice Eq.~\ref{ex4} can be satisfied if the first term in square brackets satisfies Eq.~\ref{se2} and the second expression in square brackets satisfies
\begin{equation}
V(x)=2\frac{d}{dx}\sum_{n=1}^N\psi_n(x)e^{\kappa_n x}.
\end{equation}
The thing to appreciate here is that the particular form of the scattering solutions ($\Psi(x,E)$ in Eq.~\ref{scatwf}) has built into it solutions $\psi_n(x)$ that satisfy the bound state problem, which are directly related to the form of the reflectionless potential.

\subsubsection*{Solution to Exercise 6}
The particular form of the normalization constants in Eq.~\ref{cn} arises from the following considerations.  First assume
\begin{equation}
\lim_{x\rightarrow\infty}\psi_n(x)\sim c_n^2 e^{-\kappa_n x}.
\label{asmp}
\end{equation}  
If Eq.~\ref{km} for a particular $n$ is rearranged by dividing by $c_n^2 e^{\kappa_n x}$,
\begin{equation}
\sum_{\nu=1}^{N}\left(\frac{e^{\kappa_\nu x}}{\kappa_n+\kappa_\nu}\right)\psi_\nu(x)+\frac{-\psi_n(x)e^{\kappa_n x}}{c_n^2}+1=0.
\label{km2}
\end{equation}
Now, Eq.~\ref{asmp} can be inserted, to represent the behavior of Eq.~\ref{km} as $x\rightarrow\infty$ giving
\begin{equation}
\sum_{\nu=1}^{N}\left(\frac{c_\nu^2}{\kappa_n+\kappa_\nu}\right)+e^{-2\kappa_n x}+1=0.
\label{km3}
\end{equation}
Keeping the leading terms (i.e. dropping $e^{-2\kappa_n x}$) and writing the system as a matrix equation for all $n=1, 2,...,N$ gives
\begin{equation}
{\bf Q}\cdot\vec{d}=-1
\end{equation}
where
\begin{equation}
Q_{n\nu}=\frac{1}{\kappa_n+\kappa_\nu}
\label{qnn}
\end{equation}
and $\vec{d}=\left(c_1^2, c_2^2,...,c_N^2\right)$.  So, formally,
\begin{equation}
\vec{d}=-{\bf Q}^{-1}.
\end{equation}
After some manipulation, the matrix $-{\bf Q}^{-1}$, with matrix elements given by Eq.~\ref{qnn}, gives the form Eq.~\ref{cn} for the $c_n^2$.  The sign for $c_1^2$ alone is negative using this approach, but this has no physical consequence.  Rather than finding a fixed analytical form, it is perhaps easier to write some Mathematica code to carry out the inversion of ${\bf Q}$ for arbitrary $\kappa_n$ for various fixed $N$ to observe the trend.

%\vspace{.1cm}
\begin{acknowledgements}
Thanks to Matthew Moelter, Thomas Bensky, and Katharina Gillen for their helpful feedback and discussions.  This work was performed with support from NSF grants PHY-0969852 and PHY-0855524.
\end{acknowledgements}
%\acknowledgements{Thanks to Matthew Moelter, Thomas Bensky, and Katharina Gillen for their helpful feedback and discussions.  This work was performed with support from NSF grants PHY-0969852 and PHY-0855524.}
 %\begin{enumerate}
%\item 
%\end{enumerate}


\begin{thebibliography}{99}

\bibitem{Harris07} Randy Harris, \textit{Modern Physics (2nd Edition)}, (Addison-Wesley, 2007).

\bibitem{Griffiths05} David J. Griffiths, \textit{Introduction to Quantum Mechanics, Second Edition}, (Pearson Education, Inc., New Jersey, 2005).  For the citation related to advanced courses, Chapter 11 is dedicated to an introduction to partial wave analysis and the Born approximation.  Also, there is an introduction to S-matrix and T-matrix formalism in one dimension in Chapter 2.  This might be covered in a second quarter advanced undergraduate quantum course.  For the citation related to the exercises, see Problem 2.51 on p. 89.

\bibitem{Shankar94} R. Shankar, \textit{Principles of Quantum Mechanics, 2nd Edition},  (Plenum Press, 1994).

\bibitem{Sakurai10} J. J. Sakurai, \textit{Modern Quantum Mechanics (2nd Edition)}, (Addison-Wesley, 2010).

\bibitem{Chadan77} K. Chadan and P.C. Sabatier, \textit{Inverse Problems in Quantum Scattering Theory}, Texts and Monographs in Physics (Springer-Verlag, Berlin, 1977).

\bibitem{Mackintosh12} Raymond S Mackintosh,``Inverse scattering: applications in nuclear physics,''  Scholarpedia, 7(11):12032 (2012).  \url{doi:10.4249/scholarpedia.12032}.

\bibitem{Halzen84} Frances Halzen and Alan D. Martin, \textit{Quarks and Leptons: An Introductory Course in Modern Particle Physics}, (John Wiley, 1984).

\bibitem{Griffiths08} David J. Griffiths, \textit{Introduction to Elementary Particles, Second Edition}, (Wiley-VCH, Weinheim 2008). 



\bibitem{Wernick04} Miles N. Wernick and John N. Aarsvold, Editors, \textit{Emission Tomograpy, The Fundamentals of PET and SPECT}, (Elsevier, San Diego 2004). 

\bibitem{Faddeyev63} L. D. Faddeyev and B. Seckler,  ``The inverse problem in the quantum theory of scattering,''   J.\ Math.\ Phys. \textbf{4}, 72 (1963).



\bibitem{Bimberg99} D. Bimberg, M. Grundmann, N.N. Ledentsov, \textit{Quantum Dot Heterostructures}, Vol. 471973882  (John Wiley, Chichester, 1999).

\bibitem{Jessen96} P. S. Jessen, I. H. Deutsch ``Optical lattices,'' Advances In Atomic, Molecular, and Optical Physics {\bf 37}, 95 (1996).

\bibitem{Deutsch00} I. H. Deutsch , G. K. Brennen , P. S. Jessen, ``Quantum computing with neutral atoms in an optical lattice,'' Fortsch.\ Phys. \textbf{48} (2000).

\bibitem{Nam13} D. Nam  \textit{et al.}, ``Strain-Induced Pseudoheterostructure Nanowires Confining Carriers at Room Temperature with Nanoscale-Tunable Band Profiles,'' Nano \ Lett.\ , \textbf{13}, 7, 3118Ð3123 (2013).

\bibitem{Schum08} D. Schumayer \textit{et al.},  ``Quantum mechanical potentials related to the prime numbers and Riemann zeros,''   Phys.\ Rev.\ E.\  \textbf{78}, 056215 (2008)  \url{http://arxiv.org/abs/0811.1389}.

\bibitem{Muss97} G. Mussardo,  ``The quantum mechanical potential for the prime numbers,''   \url{http://arxiv.org/abs/cond-mat/9712010}, 1997.  This work unfortunately remains unpublished, but is a very accessible and elegant application of the WKB-inspired inverse method.

\bibitem{Novikov84} S. Novikov, et al.,   \textit{Theory of Solitons},   (Consultants Bureau, New York, 1984).  In particular, see Chapter 1 sections 1-3 for an overview of inverse scattering theory related to this work.

\bibitem{Cooper95} F. Cooper, A. Khare, and U. Sukhatme, ``Supersymmetry and quantum mechanics,'' Phys.\ Rep.\  \textbf{251}, 267 (1995) \url{http://arxiv.org/abs/hep-th/9405029}.


\bibitem{Kiri98} N. Kiriushcheva and S. Kuzmin, ``Scattering of a Gaussian wave packet by a reflectionless potential,'' Am. J. Phys. \textbf{66}, 867 (1998).  \url{http://dx.doi.org/10.1119/1.2787015}.

\bibitem{Lekner07} J. Lekner, ``Reflectionless eigenstates of the sech2 potential,''  Am. J. Phys. \textbf{75}, 1151 (2007).  \url{ http://dx.doi.org/10.1119/1.2787015}.

\bibitem{Barcilon74} V. Barcilon,  ``Iterative solution of the inverse Sturm-Liouville problem,''   J.\ Math.\ Phys. \textbf{15}, 429 (1974).


\bibitem{Kay56} I. Kay and H. E. Moses,  ``Reflectionless transmission through dielectrics and scattering potentials,''   Jour.\ App.\ Phys. \textbf{27}, 1503 (1956).



\bibitem{Loudon55} R. Loudon, ``One-dimensional Hydrogen Atom,'' Am. J. Phys. \textbf{ 27}, Issue 9, pp. 649-655 (1959).  \url{http://dx.doi.org/10.1119/1.1934950}.


%\bibitem{Griffiths05} David J. Griffiths, {\em Introduction to Quantum Mechanics, Second Edition}, (Pearson Education, Inc., New Jersey, 2005).   

\bibitem{losttv} \textit{Lost} was a science fiction TV show that ran on ABC from 2001 until 2007.  Note, the numbers have no physical significance to the author's knowledge, but serve as a symbol of uncanny coincidence, often unpleasant, in the fiction of the show. \url{http://ww2.abc.go.com/shows/lost}.

\end{thebibliography}
\end{document}